\documentclass[12pt]{article}
\usepackage{amsmath}
\usepackage{graphicx}
\usepackage{float}
\usepackage{natbib}
\usepackage{url} 
\usepackage{xcolor} 
\usepackage[linkcolor=blue,citecolor=blue,urlcolor=blue,colorlinks=true]{hyperref} 
\usepackage{eurosym}

\newcommand{\blind}{0}

\usepackage{caption}
\usepackage{subcaption}
\begin{document}

\def\spacingset#1{\renewcommand{\baselinestretch}%
{#1}\small\normalsize} \spacingset{1}


\if0\blind
{
  \title{\bf Machine Learning Applied to the Detection of Mycotoxin in Food: A Review}
  \author{Alan Inglis\\
    Hamilton Institute, Maynooth University\\
     Andrew Parnell\\
    Hamilton Institute, Maynooth University\\
     Natarajan Subramani\\
     School of Biology and Environmental Science, University College Dublin\\
     Fiona Doohan\\
      School of Biology and Environmental Science,
    University College Dublin\\
    }
  \maketitle
} \fi

\if1\blind
{
  \bigskip
  \bigskip
  \bigskip
  \begin{center}
    {\LARGE\bf Title}
\end{center}
  \medskip
} \fi

\bigskip

\abstract{Mycotoxins, toxic secondary metabolites produced by certain fungi, pose significant threats to global food safety and public health. These compounds can contaminate a variety of crops, leading to economic losses and health risks to both humans and animals. Traditional lab analysis methods for mycotoxin detection can be time-consuming and may not always be suitable for large-scale screenings. However, in recent years, machine learning (ML) methods have gained popularity for use in the detection of mycotoxins and in the food safety industry in general, due to their accurate and timely predictions. We provide a systematic review on some of the recent ML applications for detecting/predicting the presence of mycotoxin on a variety of food ingredients, highlighting their advantages, challenges, and potential for future advancements. We address the need for reproducibility and transparency in ML research through open access to data and code.  An observation from our findings is the frequent lack of detailed reporting on hyperparameters in many studies as well as a lack of open source code, which raises concerns about the reproducibility and optimisation of the ML models used. The findings reveal that while the majority of studies predominantly utilised neural networks for mycotoxin detection, there was a notable diversity in the types of neural network architectures employed, with convolutional neural networks being the most popular.}

\noindent%
{\it Keywords:}  Machine Learning --- Predictive Model --- Mycotoxin --- Food Safety
\vfill

\newpage
\spacingset{1.5} 

\section{\textbf{Introduction}}
\label{sec:intro}

Mycotoxins are a group of naturally occurring, toxic chemical compounds produced by certain species of moulds (fungi),  during growth on various crops and foodstuffs, including cereals, nuts, spices and dairy products \citep{WHO}. 
The ingestion of certain mycotoxins has been linked to a range of harmful health impacts on both humans and animals, from short-term poisoning to long-term consequences such as liver cancer, and in some cases, death \citep{mavrommatis2021impact, marroquin2014mycotoxins, liu2010global}. Mycotoxins are secondary metabolites (that is, compounds produced by an organism that are not essential for its primary life processes) and are often produced during the pre-harvest, harvest, and storage phases under favourable conditions of humidity and temperature \citep{marroquin2014mycotoxins,van2022decision}. The most prevalent mycotoxins include aflatoxins, tricothecenes, fumonisins, zearalenones, ochratoxins and patulin, and are produced by certain plant-pathogenic species of \textit{Aspergillus}, \textit{Fusarium}, and \textit{Penicillium} \citep{tola2016occurrence}.  Mycotoxin contamination in crop products has been found to vary significantly across different geographical locations and is influenced by annual weather conditions \citep{logrieco2021perspectives, leggieri2020impact}. However, since 2012, there has been a noted increase in the occurrence of mycotoxins in Europe, with the impacts of climate change being most likely a contributing factor \citep{zingales2022climate, medina2017climate}. An estimated 60–80\% of the world's crop supply is contaminated by mycotoxins, and an estimated 20\% of those crops surpass the legally mandated food safety thresholds set by the European Union (EU) \citep{eskola2020worldwide}. 

With the world's food supply chain being highly interconnected, the presence of mycotoxins not only endangers human health but also has an impact on the stability of agricultural markets and trade \citep{alshannaq2017occurrence, marroquin2014mycotoxins}. The economic impact of mycotoxin contamination is substantial, with a global estimate in the billions of euro for detection, regulation enforcement, and mitigation efforts to manage mycotoxin presence in food and feeds annually \citep{wu2015global}. It is estimated that, between 2010 and 2019, approximately 75 million tonnes of wheat in Europe, which constitutes 5\% of the wheat intended for human consumption, surpassed the maximum threshold for DON contamination. This excess led to the reclassification of this contaminated wheat grain as `animal feed', resulting in an economic loss of around \euro{3 billion}  \citep{johns2022emerging}. Additionally, \cite{latham2023diverse} show that between 2010 and 2020, aflatoxins were responsible for the demotion of 4.2\% of wheat intended for food, which potentially represented an additional economic loss of \euro{2.5 billion}. As a result, the detection and management of mycotoxins in crops and food products is crucial for ensuring food safety and safeguarding consumer health worldwide, as well as contributing to economic stability.

According to  \cite{whitaker2003standardisation}, the standard methodology for mycotoxin detection comprises three main steps: sampling, sample preparation, and analytical determination. Chromatographic techniques, such as high-performance liquid chromatography (HPLC) and gas chromatography mass spectrometry (GC-MS), along with immunoassay-based methods like enzyme-linked immunosorbent assays (ELISA), are widely recognised as the most prevalent analytical approaches for the detection of mycotoxins  \citep{anfossi2016mycotoxin, maragos2004emerging}. The mycotoxin level in a bulk load is determined by measuring a sample taken from the food source. From this, the concentration of mycotoxins in the entire load is assumed to be the same as the concentration of the sample. However, these techniques often require extensive sample preparation, sophisticated equipment, and highly trained personnel, leading to significant costs and time delays in the analytical process. Furthermore, the varied and intricate nature of different foods requires customised detection methods, which can add complexity to the screening process \citep{soares2018advances, renaud2019mycotoxin}.

While traditional detection methods such as HPLC, GC-MS and ELISA generate reliable data, they often result in large, complex datasets that require extensive interpretation and analysis. Machine Learning (ML) approaches for both detection and prediction of the presence of mycotoxins have seen a rise in recent years as an alternative to traditional detection methods (see Figure \ref{fig:allpap}). At its core, ML employs statistical methods to create algorithms that allow computers to learn from data and make decisions based on identified patterns and inferences, without being explicitly programmed for each specific task. ML methods offer a sophisticated approach to deciphering the complex patterns hidden within the data and are adept at processing and analysing large datasets and extracting meaningful patterns that are not immediately apparent. By leveraging ML algorithms, researchers can gain deeper insights into the data and offer a significant advantage, when compared to traditional lab analysis, in terms of efficiency, cost, and scalability, as well as maintaining or improving the accuracy of mycotoxin detection \citep{liakos2018machine}.

ML methods can be, broadly, broken into three categories. That is, supervised learning (SL), unsupervised learning (UL), and reinforcement learning (RL). In SL, an algorithm is trained using a dataset that includes both inputs and the corresponding outputs. The model learns to associate the inputs with the outputs. After training, the model can apply this learned relationship to predict the outputs for new, unseen inputs \citep{bacstanlar2014introduction}. In UL, an algorithm is presented with only the input data and identifies patterns and structures in the data based only on the inputs. After training, it can classify new inputs based on the patterns it has found. In RL, an algorithm learns to make decisions by performing actions to achieve a goal. It processes feedback through rewards or penalties associated with its actions, using this information to develop a decision-making framework that aims to maximise rewards \citep{alpaydin2020introduction}. 

Within these categories, many different types of ML models exist and are used based on the specificity of the problem. The most popular of these models, as found by this research, are discussed in detail below. Although ML applications in food safety and mycotoxin detection are widespread, there appears to be a lack of comprehensive reviews that cover the broad spectrum of ML methodologies specifically tailored to mycotoxin analysis, as most studies tend to concentrate on individual techniques. For example, \cite{torelli2012influence} use neural networks (NN) for the prediction of contamination from the mycotoxin fumonisin in corn. Additionally, NN have been used to forecast accumulation of the trichothecene mycotoxin deoxynivalenol (DON) in barley seeds \citep{mateo2011multilayer} and to predict fungal growth \citep{panagou2009application}. For a comprehensive review of the use of NN in food science see \cite{zhou2019application}, and for a review of ML methods in general in the field of food safety, see \cite{wang2022application}, and in agriculture, see \cite{liakos2018machine}.

ML techniques can alleviate some of the current burdens of mycotoxin detection by providing an efficient and low-cost solution \citep{bernardes2022deep}. Additionally, with the impact of climate change, the need for these models to provide reliable predictions at the farm level is increasingly crucial, especially in terms of food safety and health. In this work we present a comprehensive systematic review of some of the more popular ML techniques used in the detection and prediction of mycotoxin on a range of foods and crops. Our review also identifies critical areas in the current body of work that warrant attention. A notable concern is the often insufficient discussion on the selection and tuning of hyperparameters in ML models, which is crucial for understanding and replicating study results. This lack of detail creates issues with the reproducibility of the reviewed methods and also hinders the advancement and application of these techniques.

The organisation of our article is as follows. In Section \ref{sec:searchMethod}, we provide details regarding our literature search methodology. This includes a description of the search criteria and keywords, as well as discussing the prevalence of each ML method.  In Section \ref{sec:intro2ML}, we provide a short introduction to the ML process and describe some of the common terms. In Section \ref{sec:applications} we give a brief introduction to the main ML algorithms used (and their hyperparameters) and discuss the outcomes of the articles reviewed based on the type of machine learning model used. Finally in Section \ref{sec:conclusion}, we provide some concluding remarks.


\section{\textbf{Literature Search Methodology}}
\label{sec:searchMethod}
The literature search for this review was primarily conducted using Scopus\footnote{https://www.scopus.com}, a widely recognised academic search engine that indexes scholarly articles across various disciplines. To ensure the relevance of the research, the search was restricted to articles published within the last 10 years (since November 2023). This time frame was chosen to capture the most recent advances and trends in the application of machine learning to mycotoxin detection in crops. The search engine was used to identify key studies, reviews, and seminal works pertinent to the topic at hand.

\subsection{\textbf{Search Criteria and Overview}}
A comprehensive search was conducted on the Scopus database and focused on publications between the years 2013 to 2023. The search was conducted using the primary keyword \textit{``mycotoxin''} in combination with machine learning-related terms: \textit{``artificial intelligence''}, \textit{``bagging''}, \textit{``bayesian network''}, \textit{``boosting''}, \textit{``decision tree''}, \textit{``deep learning''}, \textit{``ensemble''}, \textit{``gradient boost''}, \textit{``k-means''}, \textit{``k-nearest neighbour''}, \textit{``knn''}, \textit{``machine learning''}, \textit{``neural network''}, \textit{``principal component analysis''}, \textit{``random forest''}, \textit{``supervised learning''}, \textit{``support vector machine''}, \textit{``SVM''}, and \textit{``unsupervised learning''}. The search terms were motivated by a similar search used in a review of machine learning to the monitoring and prediction of food safety by \cite{wang2022application}. This strategy was employed to ensure a wide coverage of potential articles at the intersection of mycotoxin detection and machine learning methodologies.

\begin{figure}[H]
    \centering
    \includegraphics[width=0.6\textwidth]{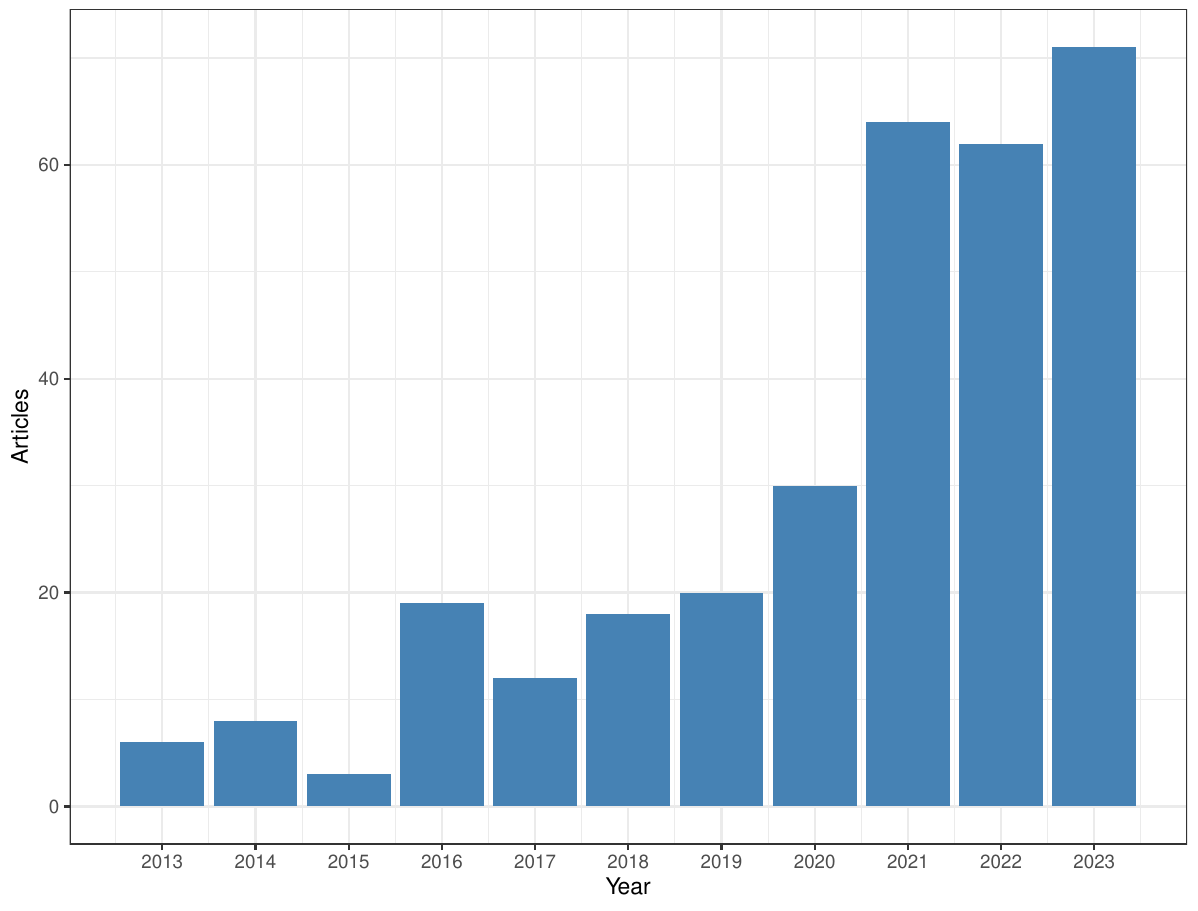}
    \caption{Number of publications between 2013 and 2023 found by our systematic search criteria in Scopus.}
    \label{fig:allpap}
\end{figure}

This search yielded 313 documents on Scopus. Figure \ref{fig:allpap} shows the results obtained from Scopus over the years 2013 to 2023. There is a general increasing trend across the years, with a marked rise after the year 2021.

\begin{figure}[H]
    \centering
    \includegraphics[width=0.7\textwidth]{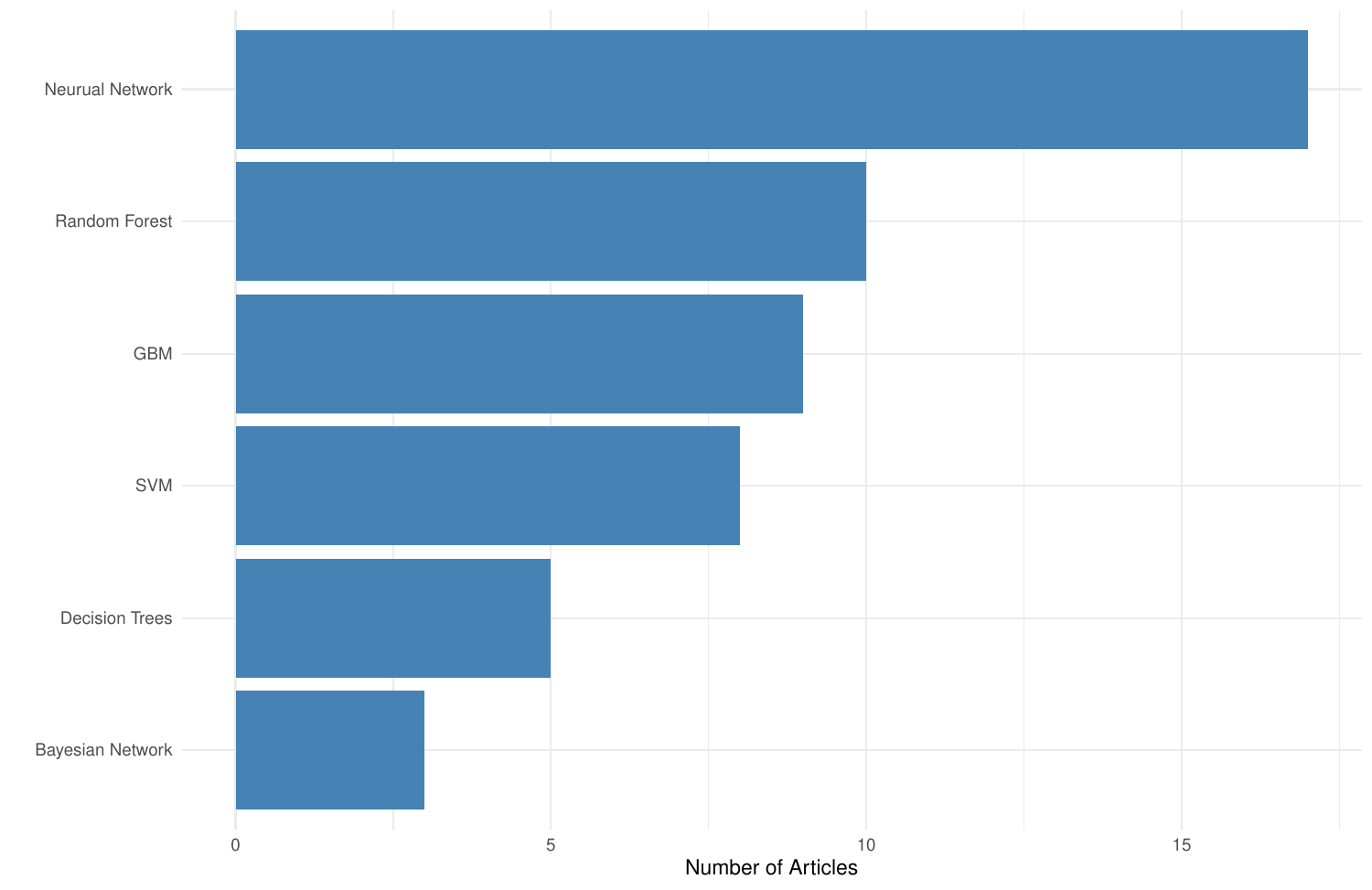}
    \caption{Most popular machine learning methods reviewed in this work.}
    \label{fig:mlmethod}
\end{figure}

To limit the search further, only peer reviewed articles in English, in the fields of agricultural and biological sciences, environmental science, computer science, and mathematics were chosen. This reduced to search size to 91. After examining the abstracts of all 91 articles, 45 were selected for their relevance and included in this study. From these articles, the predominant ML technique used was neural networks (NN), followed by random forests (RF) and gradient boosting (GB), then support vector machines (SVM), decision trees (DT), and Bayesian networks (BN). Figure \ref{fig:mlmethod} shows the frequency of each ML algorithm used in the literature.
\\
\\
\\

\section{\textbf{A Brief Introduction to Machine Learning}}
\label{sec:intro2ML}

In this section we provide a general overview of the ML process. This fore-knowledge is useful when discussing the ML approaches reviewed later in this document, though those already with experience in this topic may skip this section. To begin, we describe the typical process of creating an ML model.

\subsection{Typical Machine Learning Process}

\begin{figure}[H]
    \centering
    \includegraphics[width=\textwidth]{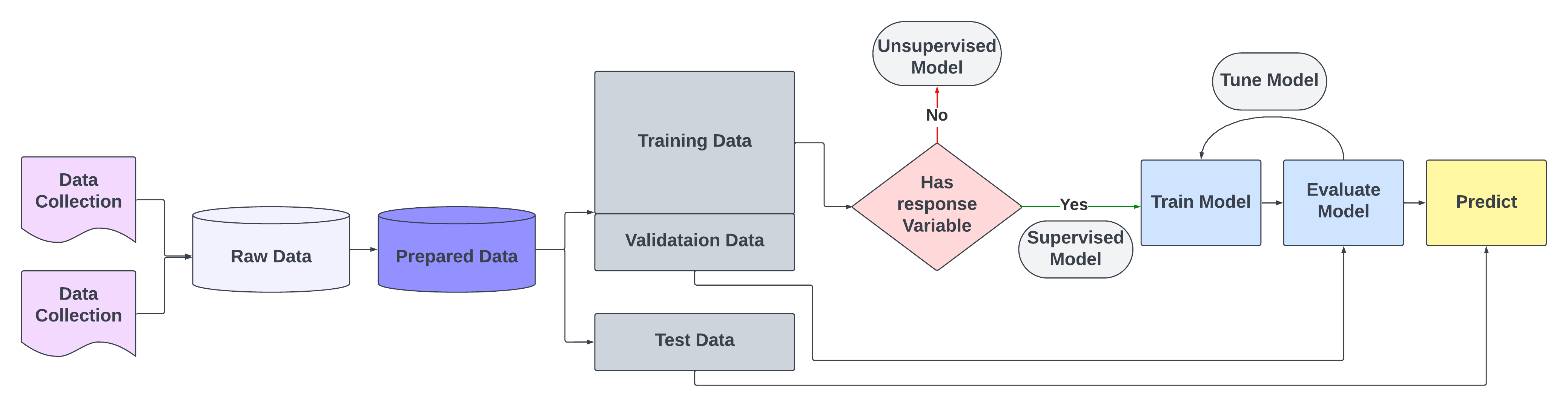}
    \caption{Typical machine learning process.}
    \label{fig:MLprocess}
\end{figure}

Figure \ref{fig:MLprocess} shows a typical ML process for unsupervised and supervised learning methods.

We can break up the process outlined in Figure \ref{fig:MLprocess} into five distinct steps. These are, 

\begin{enumerate}
    \item \textbf{Data Collection:} The process starts with the collection of raw data, which can be from many sources or sites.
    \item \textbf{Data Preparation:} This raw data is then prepared for analysis. This process typically involves cleaning and formatting the data.
    \item \textbf{Data Splitting:} After preparation, the data can be split into three parts. These are; training data, validation data, test data (discussed more below).
    \item \textbf{Model Selection:} Depending on the type of data, either an unsupervised or supervised learning model (or models) is chosen.
    \item \textbf{Model Training, Evaluation, and Prediction:} This process involves training the model with training data, optimising the hyperparameters of the model using the validation data, and then evaluating its overall performance using test data.
\end{enumerate}

\subsection{Training, Validation, and Test Data}
In ML, validation and test data are crucial for developing and evaluating models. Validation data is a separate subset of the original data, not used in training the model (see Figure \ref{fig:MLprocess}).  It helps in fine-tuning the model's parameters (known as hyperparameters) which are pre-set configurations of the model. This fine-tuning of hyperparameters during the validation process is essential to optimise the model's performance. Validation also assists in selecting the best version of the model by providing feedback on its performance. This step is essential to prevent overfitting, ensuring that the model doesn't just memorise the training data but learns to generalise from it, making accurate predictions on new, unseen data. Test data is used after the model has been trained and validated (see Figure \ref{fig:MLprocess}). It is another distinct subset of the data set, not used in either training or validation. The test data is used to evaluate the final model's performance, providing an unbiased assessment of how well the model is likely to perform in real-world scenarios. 

In all the referenced studies we cover below the model performance is quantified by evaluating the model performance on the test data set, unless otherwise stated. Sometimes authors also report the training or validation data set performance but, for the reasons outlined above, these should be discarded as a measure of model performance. The common performance metrics used in these studies include:
\begin{itemize}
    \item $R^2$: This statistic measures the proportion of variance in the dependent variable that can be explained by the independent variables in the model. An $R^2$ value closer to 1 indicates that the model accounts for a significant amount of the variance in the dependent variable.
    \item MSE and RMSE: Mean Square Error (MSE) is the average of the squares of the errors, which are the differences between predicted and actual values. Lower MSE values indicate a better fit of the model to the data. Root Mean Square Error (RMSE) is the square root of MSE. It has the same units as the quantity being estimated (for regresion problems) and provides a measure of the differences between model's predicted values and the actual observed values. Like MSE, a lower RMSE is better.
    \item Accuracy: This metric is commonly used for classification tasks and represents the ratio of correctly predicted observations to the total observations. High accuracy indicates that the model can correctly classify the instances with high reliability.
    \item AUC: Area Under the Receiver Operating Characteristic Curve (AUC) is used in binary classification to measure a model's ability to distinguish between classes. An AUC of 1 represents perfect classifier performance, while an AUC of 0.5 denotes a model with no discriminative power.
\end{itemize}


\section{\textbf{Application of Machine Learning to Mycotoxin Data}}
\label{sec:applications}
In this section we discuss the most common ML algorithms (from Figure \ref{fig:mlmethod}) and review their application to mycotoxin data. Each subsection is dedicated to a single ML method in which we describe the basic algorithm, how it makes predictions/detections, some advantages and disadvantages of the algorithm, and finally a review of literature using these methods. In cases where the reviewed studies employ multiple machine learning models, we categorise each paper based on the highest-performing model used in that particular work.

\subsection{\textbf{Neural Networks}}
\begin{figure}[H]
    \centering
    \includegraphics[width=0.6\textwidth]{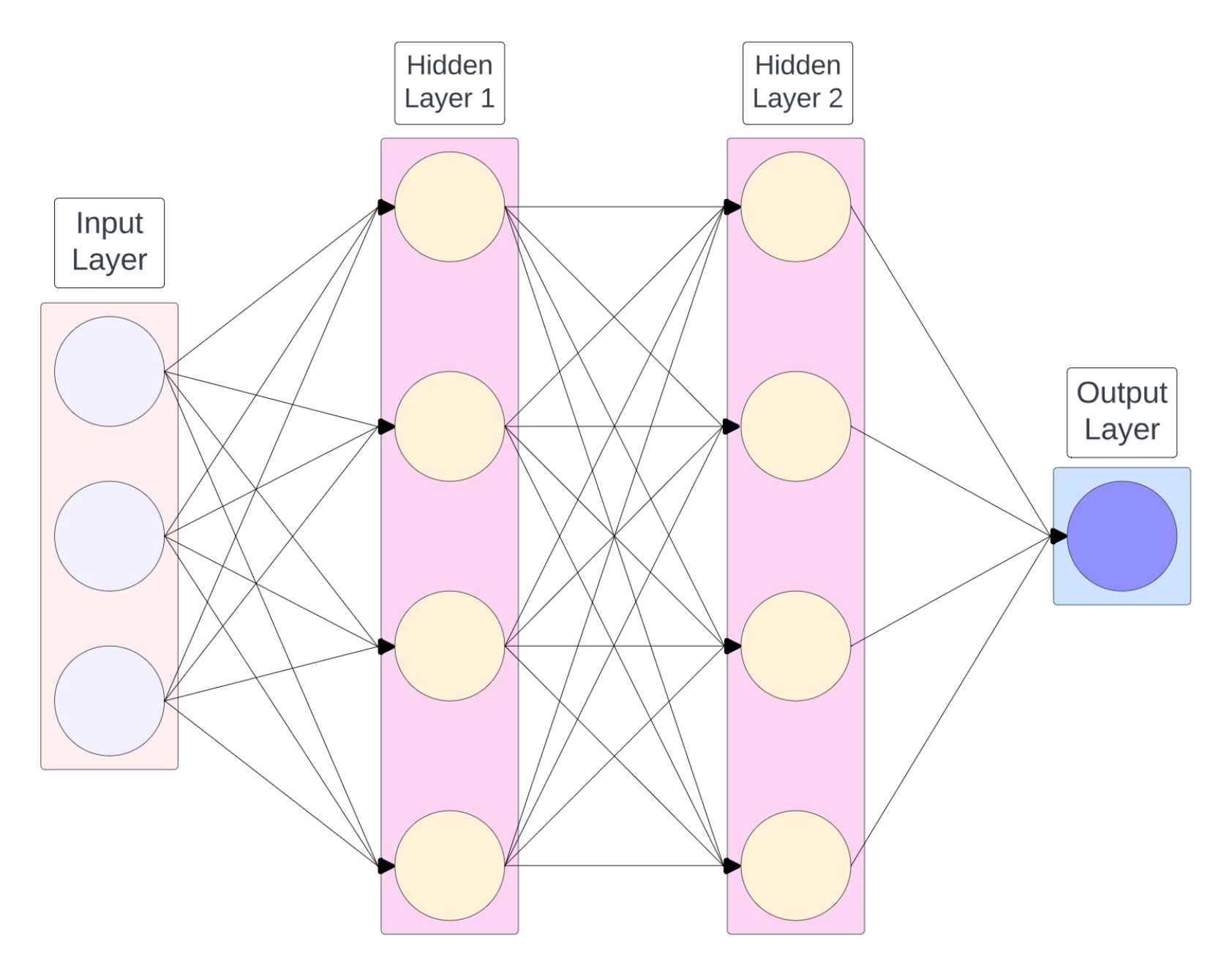}
    \caption{Basic neural network structure, showing an input layer, two hidden layers, and an output layer, where each circle represents a neuron and are interconnected by lines symbolising neural connections. The input layer receives the initial data, which is then processed through successive hidden layers using weights and activation functions, refining the information before it reaches the output layer.}
    \label{fig:nnet}
\end{figure}

Neural networks (NNs), first introduced by \cite{mcculloch1943logical}, are a class of machine learning algorithms modelled loosely after the human brain \citep{gurney2018introduction}. They are designed to identify patterns and make predictions by learning from data and can be used for supervised or unsupervised problems. NNs are made up of interconnected nodes and edges, where the nodes represent the \textit{neurons} and the edges are the links between the neurons. The nodes are organised into layers, where the first layer is called the input layer, the last layer is the output layer, and all intermediate layers are called hidden layers. Typically, in an NN, the data is fed to the input layer, then one or more hidden layers perform computations and learn from the data, and finally predictions (or classifications) are provided by the output layer. A simple diagram of an NN can be seen in Figure \ref{fig:nnet}.

Every neuron in a hidden layer applies a weighted sum of the inputs to transform the data. This is followed by a function, referred to as an \textit{activation} function \citep{gurney2018introduction}. The network fine-tunes the weights associated with each neuron by employing optimisation algorithms throughout the training phase. There are numerous hyperparameters associated with NNs. Some of the main hyperparameters include: (i) the learning rate which determines how much the weights are changed at each iteration; (ii) the number of epochs, which refers to how many times the entire training dataset is passed forward and backward through the neural network;; (iii) the batch size which controls the number of training examples used in one iteration; and (iv) activation functions like ReLU (Rectified Linear Unit), sigmoid, and tanh that determine the output value of a node given an input or set of inputs. After training, the NN is capable of generating predictions for new, unseen data by passing the input across the layers to produce an output. 

Like all machine learning models, NNs come with their own set of advantages and disadvantages. For example, NNs excel at identifying and modelling non-linear interactions present in data, which are common in biological processes. They are also flexible and can handle a wide range of data types, such as numerical and categorical, text, and image data. Despite their advantages, neural networks also have limitations. One of the major limitations is interpretability. NNs are considered \textit{black-box} algorithms, meaning that it is difficult to understand why specific predictions are being made \citep{montavon2018methods}. Secondly, like many of the other ML approaches we cover, they are not probabilistic models, making it hard to accurately quantify the uncertainty in the predictions.  Overfitting can also be an issue for NNs. Without appropriate regularisation, NNs can become too complex, capturing the noise in the training data instead of generalising to the underlying pattern \citep{srivastava2014dropout}. Finally, training large NNs requires a significant amount of computing power. The computational cost of NNs will increase with the complexity of the model \citep{canziani2016analysis}. In the following subsections, we review at the use of NN on different types of mycotoxin data. 

\subsubsection{NNs applied to Spatiotemporal Data}

NNs have been widely applied to spatio-temporal data, despite them not forming part of the traditional suite of spatio-temporal analytics techniques. In the field of mycotoxin study, NNs have been used for a variety of tasks and data types. For example, \cite{camardo2021machine} use data from several sites in Northern Italy over the years 2005 to 2018. Their goal was to predict the presence of mycotoxins (specifically, aflatoxin and fumonisins) using NNs in corn. In their work they trained two NNs to predict if the contamination levels were above legal thresholds at the time of harvest. Both models performed well, achieving an accuracy of greater than 75\% on the test data. However, they recommend, for future research, that improvements can be made to the modelling by taking into account the  co-occurrence of aflatoxin and fumonisins in corn and their complex interaction, which may be due to the effects of climate change.

\cite{niedbala2020application} applied NNs to analyse the concentration of mycotoxins in winter wheat grain. They examined 23 winter wheat genotypes with different Fusarium resistance, from three different sites in Poland during the years 2011 to 2013. They developed three NN models, however only two of these are concerned with the detection of mycotoxins, that is the DONANN model which is used to detect DON and the NIVANN model, which examines the nivalenol content. The DONANN and NIVANN models were designed using an automatic network designer using \texttt{Statistica v7.1} software \citep{StatSoft}, and were evaluated among a set of 10,000 generated networks. The performance of these models was assessed on several statistical metrics, but the primary focus was on the correlation coefficient (which, in this case, would be the correlation between the predicted values from the model and the actual observed values) and the mean absolute error (MAE), which is the absolute differences between the predicted values and the actual values. For the best performing DONANN model, a low MAE of 0.37 was reported, however, the correlation coefficient was exceptionally high at 0.99, indicating an almost perfect linear relationship between the predicted and actual values. The  best performing NIVANN model, while exhibiting a slightly lower correlation coefficient of 0.81 and an MAE of 0.02, still performed within acceptable ranges. The architecture of the created models was designed as a multi-layer perceptron (MLP) type of NN, with two hidden layers. Despite reporting of training, validation, and test errors, the authors did not specify the data set on which the correlation and MAE metrics were based.

In a novel application of NNs, \cite{jubair2021gptransformer} use a transformer-based deep learning method, called \textit{GPTransformer}. A transformer-based deep learning algorithm refers to a type of NN architecture that relies on a mechanism called \textit{attention} to boost the performance of the model \citep{vaswani2017attention}. In their work, the authors propose a transformer-based genomic prediction model for predicting Fusarium head blight disease levels and associated Fusarium DON concentration in barley data collected in three locations in Canada over the years 2014 to 2015. One of their goals was to compare the accuracy of the GPTransformer model to existing genomic prediction methods such as decision tree algorithms (DT), linear regression (LReg), and traditional statistical algorithms like best linear unbiased prediction (BLUP). The authors use the Pearson correlation coefficient (PCC) as a measure of performance which calculates the linear relation between the true output and predicted output. They show that the GPTransformer model (and all of the used ML models) did not significantly outperform the statistical method of BLUP in terms of predictive accuracy. However GPTransformer did perform better than both the DT and LReg methods. The authors note that the ML methods used are able to capture non-additive genetic elements and as such the predictions provided might include some of these interactions in their estimations. 

\subsubsection{NNs Applied to Spectral Data}
Hyperspectral (or just spectral) data refers to capture and processing of information from across the electromagnetic spectrum \citep{grahn2007techniques}. \cite{jin2018classifying}, \citet{qiu2019detection}, and \cite{rangarajan2022detection},  all apply NN classification algorithms to pixels of hyperspectral image data to examine wheat for Fusarium head blight infection. Each author uses a convolutional NN (CNN), which captures spatial patterns or motifs by identifying and calculating weights from the images according to how often the motif appears. 

In \cite{rangarajan2022detection} the authors investigated four distinct methods for converting hyperspectral imaging data. They then evaluated the performance of eight different CNN models in classifying pixels as either healthy or infected with Fusarium Head Blight. The effectiveness of these models was compared based on their classification accuracy. They found that a particular type of CNN called \textit{DarkNet 19} \citep{darknet} performed the best, with an accuracy of close to 100\% across all data conversion methods, on both the validation and test data. For \cite{qiu2019detection}, tests showed that the CNN model is effective in detecting images that contain the blight and achieved an $R^2$ value of 0.80  and the mean average accuracy for the testing data set was 92\%. In \cite{jin2018classifying}, the authors compared the accuracy of the different NNs to determine which is the best at identifying diseased regions of the wheat kernel. They showed that a two-dimensional convolutional bidirectional gated recurrent unit NN performed the best, with an accuracy of 84.6\% on the validation data set and an F1 score and accuracy of 0.75 and 74.3\%, respectively, on the test data. 

\cite{han2019pixel} use a combination of hyperspectral data and NNs to detect aflatoxin in peanuts. They showed the CNNs efficacy in classifying infected peanuts and achieve a test set accuracy of 95\%. They later expanded their work and use a one dimensional CNN (1D-CNN) to classify aflatoxin infection in corn and peanuts. This time they achieved an accuracy of 96.4\% for peanuts and 92.1\% for corn \citep{gao2021aflatoxin}.

In research conducted by \citep{oener2019machine}, infrared (IR) spectroscopy and ML algorithms were used to detect fungal contamination in corn. In their study, 183 naturally infected samples (contaminated with different \textit{Fusarium} DON species and at different concentraions) were obtained from the seed production Linz of Austria (SBL), and from the Cereal Research Centre of Hungary (CRC). The authors assess several classification ML models, including multi-layer perceptron (MLP) neural networks, Random Forests, Support Vector Machines, and Adaptive boosting, for their accuracy in correctly classifying contaminated from non-contaminated samples. Their results show that the MLP approach correctly classified 94\% of the non-contaminated samples and 91\% of the contaminated samples. The authors note that while this approach yields promising results, these findings are specific to a contamination threshold of 1,250 mg/kg, which is the EU regulatory limit, and that subsequent research will aim to evaluate the performance of the classification methods across various contamination levels.

\subsubsection{NNs With an Electronic Nose}

An electronic nose (e-nose) is a device intended to detect chemical compounds in gasses. E-noses have been extensively used in the detection of aflatoxins \citep{ottoboni2018combining, campagnoli2009application}, fumonisins \citep{gobbi2011electronic}, and DON \citep{lippolis2014screening} in corn. However, \cite{leggieri2021electronic} use an e-nose supported by NNs for the detection of aflatoxin and fumonisins in corn. In their work, they compared three different approaches, that is, NN, logistic regression (LR), and discriminant analysis (DA) to examine the e-nose's ability to discriminate between samples contaminated with concentrations either exceeding or falling below legal thresholds on data spanning five years. They showed that all methodologies achieve an accuracy of above 70\%, with the NN performing the best with an accuracy of 78\% for aflatoxin detection and 77\% for fumonisin detection. They go on to suggest that the e-nose, when supported by a NN, can provide a fast screening tool for classifying samples.

\subsubsection{NN Summary}
Neural Networks have been widely adopted as the ML algorithm of choice for analysing mycotoxin data, especially in the field of hyperspectral imaging. However, as of yet, there seems to be a gap between research applications and the wider use in industry. The application of NNs in hyperspectral data for mycotoxin detection (and food safety in general) is a relatively new process and the implementation of an NN approach to hyperspectral data in industrial quality control faces various challenges, mainly due to hardware limitations, such as the cost of operating imaging equipment \citep{saha2021machine}. However, in research, NNs for use in hyperspectral imaging have seen an increase in popularity with many of the reviewed works being widely cited \citep[for example,][]{jin2018classifying, qiu2019detection}.


\subsection{\textbf{Random Forests}}
\label{sec:RF}
A random forest \citep[RF;][]{breiman2001random} is an ensemble learning method used for classification and regression. The RF algorithm creates a \textit{forest} of decision trees, where each tree in the forest is built from a sample drawn with replacement (that is, a bootstrap sample) from the training set and selects splits from a random subset of features. 

\begin{figure}[H]
    \centering
    \includegraphics[width=0.8\textwidth]{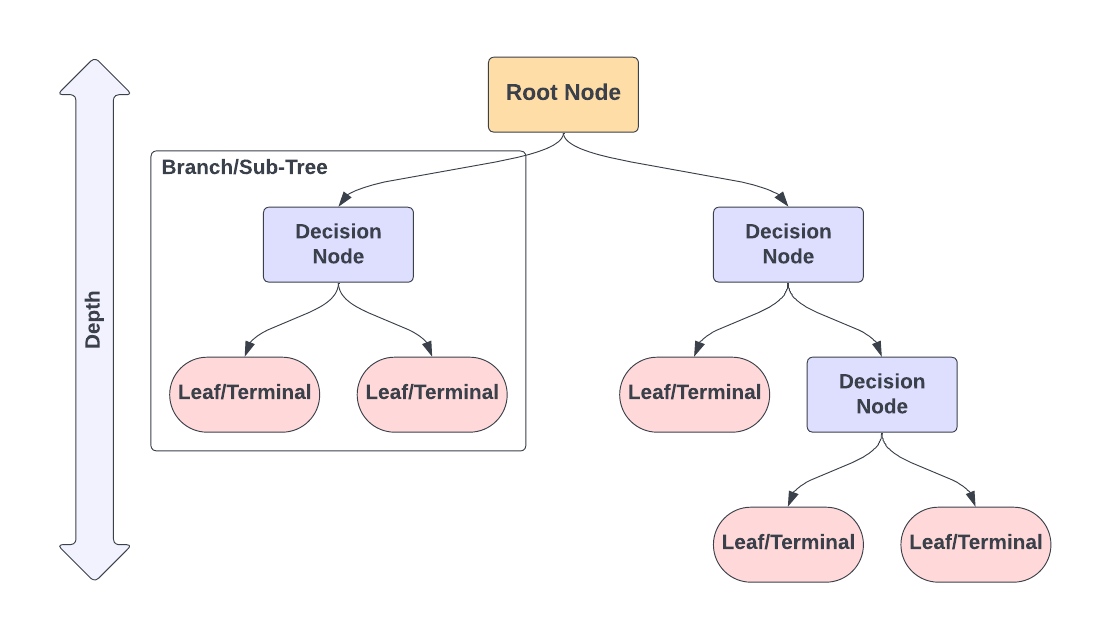}
    \caption{Decision tree process demonstrating the structure of a decision tree, including the root node, branching to decision nodes, and culminating in leaf/terminal nodes. The depth of the tree is indicated, showing the levels of decision-making from the root to the leaves.}
    \label{fig:decisionTreeSingle}
\end{figure}

While Section \ref{sec:DTs} provides a comprehensive examination of decision trees, this Section offers a concise introduction to familiarise readers with the basic concepts and terminology associated with decision trees. Figure \ref{fig:decisionTreeSingle} shows an example of a single decision tree. In constructing each decision tree, the root node is the starting point and it represents the entire dataset, which gets split based on a feature that provides the best separation according to a certain criterion \citep[like Gini impurity][]{breiman1984classification}. The decision nodes are the points where the data is split further. Each decision node represents a decision rule on a specific feature. The process continues recursively until a stopping criterion is met, such as reaching the tree's maximum depth, attaining a minimum sample count in a leaf, or achieving adequate purity within the leaf nodes. The leaf/terminal nodes represent the final output of the decision process. Each branch/sub-tree represents a possible outcome of the decision made at the decision node, leading to further sub-trees or leaf nodes.

\begin{figure}[H]
    \centering
    \includegraphics[width=0.8\textwidth]{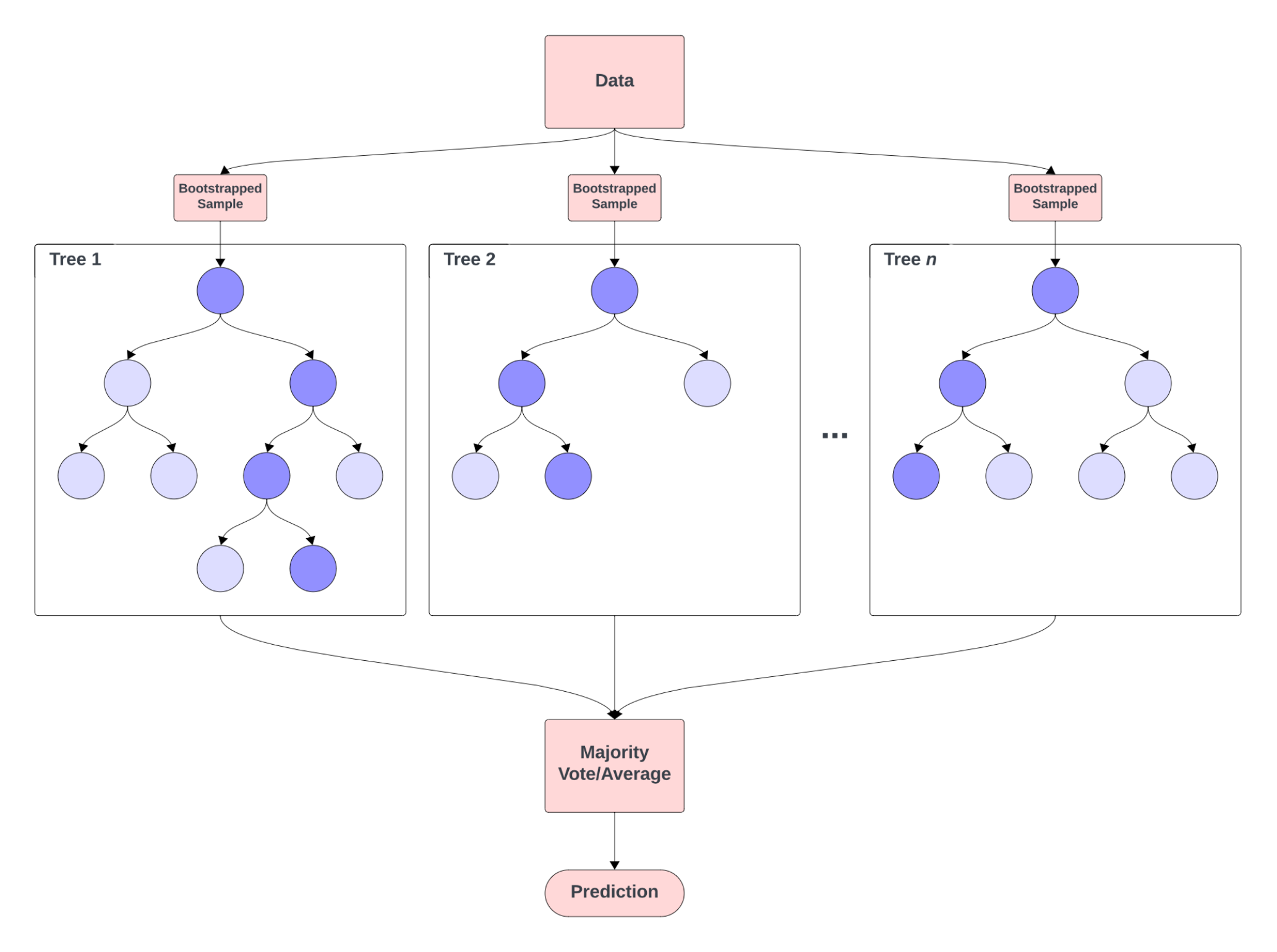}
    \caption{The Random Forest algorithm constructs an ensemble of decision trees, with each tree built from a unique bootstrapped sample of the original dataset. Distinct paths through each tree are shown, highlighted by the darker blue nodes, and represent a sequence of decisions made from the root to a leaf node based on the input features. The final prediction of the Random Forest is determined by aggregating the predictions of all trees, using majority voting for classification tasks or mean prediction for regression tasks.}
    \label{fig:rf_pic}
\end{figure}

For RF classification tasks, each tree in the forest votes for a class, and the class receiving the majority of votes becomes the model’s prediction. For regression tasks, the forest takes the average of the outputs by individual trees. Figure \ref{fig:rf_pic} shows a summary of the RF algorithm.

One of the main advantages of using RFs is their versatility. They are capable of performing both regression and classification tasks, as well as handling large datasets. Additionally, they require very little tuning and can perform well without much hyperparameter optimisation. Some of the main hyperparameters associated with RF include (i) The number of trees: this is the number of trees in the forest. Generally, more trees increase performance but also increase computational cost. (ii) Maximum depth of trees: the maximum depth of each tree. Deeper trees can model more complex patterns but might lead to overfitting. (iii) Minimum samples split: the smallest number of samples needed to split an internal node. Setting higher values helps prevent the model from learning overly specific patterns, which can lead to overfitting. As with NNs, RFs are a black-box algorithm, and so interpretability can be an issue. Each decision tree upon which the RF is built can be easy to interpret, but since RFs consist of a large number of decision trees averaged together, the decision process by which a prediction is made can be somewhat opaque.

\subsubsection{RFs for Spectral Data}
As with NNs, RFs have been applied to hyperspectral data. For example, \cite{ghilardelli2022preliminary} use a RF classification model to classify corn silage for high or low mycotoxin contamination using near-infrared spectroscopy (NIR). In their study, 155 samples were collected from several sites in the Po Valley (Italy) and from Sardinia over the years 2017 to 2019. Their aim was to develop qualitative models capable of distinguishing corn silage based on either the total concentrations or the total counts of various groups of mycotoxins (in this case, \textit{Fusarium} and \textit{Penicillium} toxins). To evaluate various classification strategies, different distinct threshold levels were established for each mycotoxin contamination. These thresholds were used to categorise each sample as having either a high or low contamination level in relation to these specified values. To predict the contamination level, an RF classification model was fitted, using the wavelength of light as the predictors, and achieved an out of sample accuracy of above 90\% for the classification of both Fusarium and penicillium toxins.

In a 2023 study, \cite{teixido2023quantification} utilised NIR spectroscopy for detecting DON in oat samples from Spain and Sweden collected over the years 2021-2022. The authors apply two different transformation techniques to the spectral data and examine which allows for greater classification of the data using four different ML algorithms (k-nearest neighbours, Naïve Bayes, NN, and RF). Both preprocessing transformation methods achieved similar results for all ML methods, with RFs performing the best with an accuracy of 77.8\% and an area under the curve (AUC) of around 0.77. However, they note that other similar studies have been conducted that achieve a higher classification accuracy, such as \cite{femenias2021near}. 

In a similar study, \cite{ma2023accurate} constructed a biosensor array for identifying mycotoxins in peanuts and corn, produced by Aspergillus flavus, using six ML models including partial least square determination analysis (sPLS-DA); linear support vector machine (svmLinear); radial support vector machine (svmRadial); RF; NN; and high dimensional discriminant analysis (HDDA). The authors use the classification models for three separate purposes: to distinguish healthy from infected samples; to distinguish the pre-mould status in infected samples; and to distinguish between infected peanuts or corn samples. To distinguish the pre-mould status, the aim was to create a three class model to predict either the control, or one-or-two days post-inoculation. Their approach achieved a reported 100\% accuracy in distinguishing healthy from infected samples, and an RF accuracy of 95\% and 98\% in identifying pre-mould status in peanuts and corn, respectively. However, such high levels of accuracy warrant further investigation, as such high accuracy rates can often be indicative of issues in the experimental design, such as the creation of non-representative test sets or overfitting, especially if the test sets are not properly randomised.

\subsubsection{RFs for Mycotoxin Treatment}

ML models in mycotoxin treatment can be used to predict mycotoxin contamination risk and optimise mitigation strategies. This application can boost accuracy in prediction and effectiveness in deploying targeted anti-fungal treatments. In a study conducted by \cite{tarazona2021machine}, the authors employ machine learning techniques to predict the growth of \textit{Fusarium culmorum} and \textit{Fusarium proliferatum}, as well as their production of mycotoxins, in environments where ethylene-vinyl alcohol copolymer films are used. These films contain pure components of essential oils, which are used to inhibit the growth of the fungi and their mycotoxin production. In their work they studied fungal growth on corn in vitro and modelled the fungal growth and toxin production under different environmental scenarios and with different treatments applied. The ML models used were NNs, RF, extreme gradient boosted trees (XGB), and multiple linear regression (MLR). The performance of the ML methods was assessed using the root mean square error (RMSE). It was found that RF performed the best in predicting the growth rate of \textit{Fusarium culmorum} and \textit{Fusarium proliferatum} and mycotoxin  production, having consistently the lowest RMSE value.

\cite{mateo2023exploring} evaluated the anti-fungal properties of specific lactic acid bacteria strains against \textit{Fusarium} species found in cereals. To achieve this, various machine learning algorithms, including  NN, RF, XGB, and MLR, were employed to predict the extent of fungal growth inhibition resulting from the application of the tested lactic acid bacteria strains. As with the previous study, the RMSE was the metric used to assess performance of the model, in conjunction with the $R^2$ value. In this work, both RF and XGB showed comparable performance, reporting similar RMSE (0.0604 and 0.0581, respectively) and $R^2$ values (0.992 and 0.992, respectively) on the test data, in predicting the percentage of growth inhibition.

Several other studies exist on the topic of using ML models (and specifically RF) to predict mycotoxin growth in the presence of treatments. In the interest of brevity and space, we name them here but do not provide additional details of the studies. In each of these studies, the authors use multiple ML models, with a general consensus that RF models perform the best at their given tasks. See \cite{mateo2021comparative, tarazona2021potential, srinivasan2022predicting} for more details.

\subsubsection{Random Forest Summary}

RFs have emerged as a robust and versatile tool in the field of mycotoxin detection and treatment and have gained popularity due to their ease of use, computational speed, and predictive performance. These studies collectively underline the significant potential of RF in enhancing food safety measures, although it is crucial to acknowledge the necessity for rigorous validation and testing to ensure the reliability of these models. 


\subsection{\textbf{Gradient Boosting }}
\begin{figure}[H]
    \centering
    \includegraphics[width=0.8\textwidth]{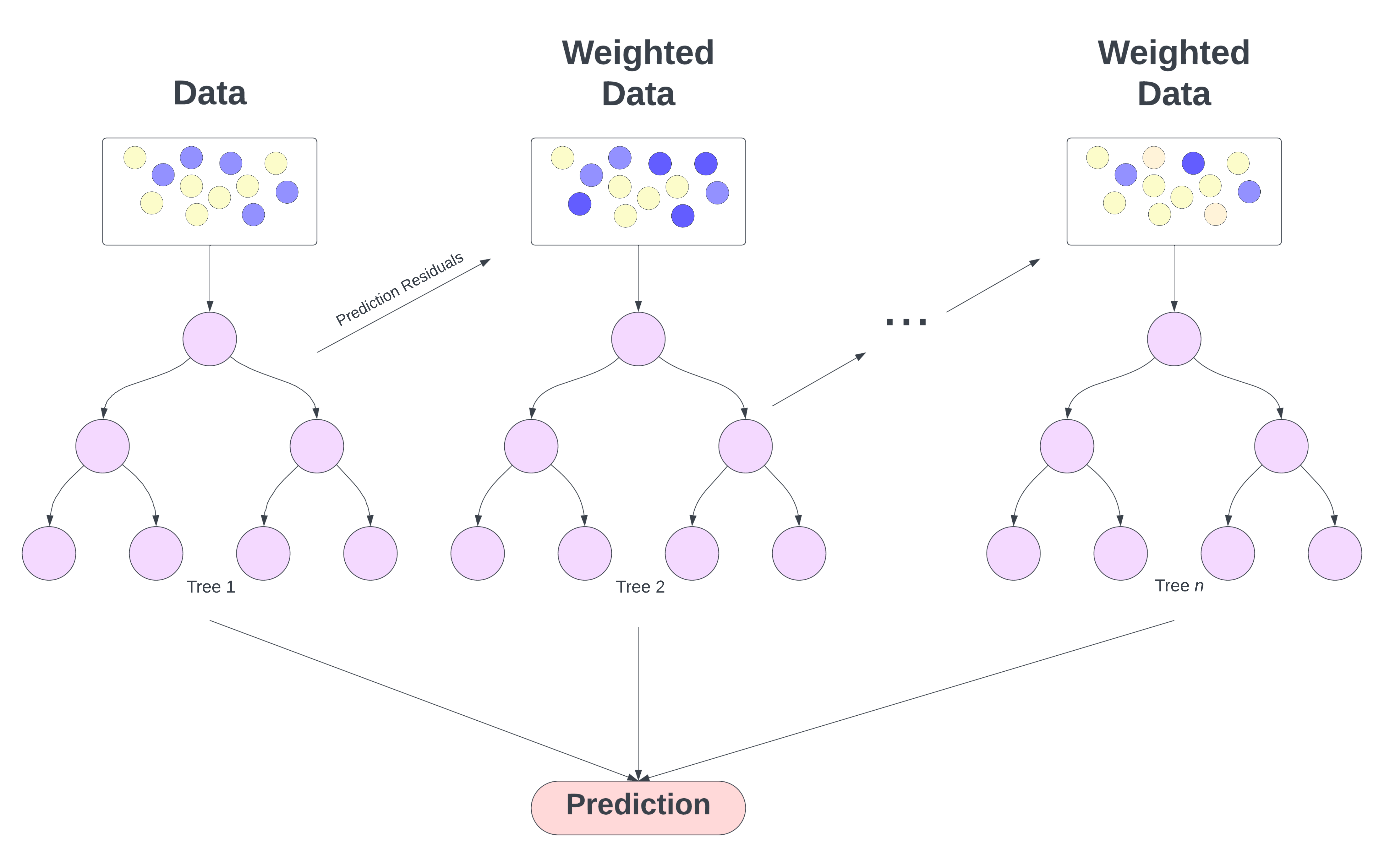}
    \caption{Gradient Boosting Process. Here the weak learners are trees that are trained sequentially on weighted data with iteratively adjusted weights based on previous prediction errors.}
    \label{fig:gbmProcess}
\end{figure}

Gradient Boosting \citep[GB;][]{friedman2001greedy} builds on the concept of boosting, where weak learners are converted into strong ones through an iterative process. The GB framework builds boosted regression models by sequentially training a weak classifier (such as a linear regression or simple decision tree) successively on the data using the residuals from previous model fits (as shown in Figure \ref{fig:gbmProcess}). This process ensures that each new weak classifier addresses the inaccuracies of its predecessors, thereby enhancing the prediction accuracy. The final model aggregates the outputs from all these weak classifiers to form a robust, `strong' classifier through an ensemble approach. The term \textit{gradient} in Gradient Boosting refers to the method's use of gradient descent, a numerical optimisation algorithm, to minimise the loss, or the difference between the actual and predicted values.

In Gradient Boosting, when the weak learners are decision trees, each tree is grown in a greedy manner, but unlike Random Forests, trees are grown sequentially. After the first tree is built and predictions are made, the errors (residuals) from those predictions are used to build the next tree. The subsequent tree aims to predict the residuals from the previous tree. This process is continued, with each new tree correcting the residuals of the ensemble of all previous trees. The final prediction is made by summing the predictions from all trees, which can be thought of as a weighted vote where trees that reduce the error the most have more influence.

An advantage of GB models is their strong predictive capability and adaptability, especially in dealing with complex, non-linear relationships between independent variables and the dependent variable. They adapt to various prediction problems by supporting different loss functions, making them suitable for both regression and classification tasks. However, these models have their challenges. Without careful tuning and regularisation, there is a risk of overfitting, a problem exacerbated by noisy data \citep{natekin2013gradient}. Additionally, their sequential boosting process is computationally intensive and time-consuming compared to methods like Random Forests that build trees in parallel. This complexity can be a significant drawback in scenarios where computational resources or time are limited. Some of the main hyperparameters associated with GB are; (i) Number of weak learners: this defines the number of boosting stages or learners to be created. More learners can lead to a more powerful model, but also increase the risk of overfitting and raise computational cost. (ii) Learning Rate: this parameter scales the contribution of each learner. A smaller learning rate requires more weak learners but can yield a more generalised model. In the case of the weak learner being trees,(iii) Maximum Depth of Trees: determines the maximum depth of each individual tree. Deeper trees can model more complex patterns but can also lead to overfitting. An extension of a GBM model is called eXtreme Gradient Boosting (XGB) \citep{chen2016xgboost}, with the key difference between the two being performance. In general, XGB models are faster and have better optimisation. Additionally, XGB models have the ability to deal with missing values.

\subsubsection{GB for Spatio-Temporal Data}
In a study by \cite{wang2022designing}, the authors design a program for aflatoxin monitoring in feed products (peanuts and soy beans), whilst considering both the performance of the model and the cost of monitoring. In the study, they apply four different ML algorithms (namely, GB, LR, SVM, and DT) to historical data concerning monitoring for the presence of aflatoxins in feed products. The data was collected from several sites around the world, including China, Brazil, and Argentina, over the years 2005 to 2018. The ML algorithms were compared to predict which feed batches are high risk and which should be considered for further aflatoxin analysis. In their work, they found that all the ML models performed well, and used several error metrics to assess their models. They obtained an accuracy of over 90\% for all models, and an AUC and recall of over 0.8 and 0.6, respectively. However, the XGB model performed better than all other models and the authors propose a reduction to the monitoring cost of up to 96\% for the years 2016 to 2018.

In \cite{xie2022fungi}, the authors propose to use un-targeted metabolomics and ML techniques to mine biomarkers of the species \textit{Aspergillus} on peanut data collected from several sites in China over the years 2013 to 2018. They initially use an RF model to determine \textit{Aspergillus} species with 97.8\% accuracy. They then go on to use XGB to create a decision rule to help regulators in evaluating risk prioritisation with a claimed accuracy of 87.2\%. However, the authors note that they build the XGB model using only a single tree and use this tree to create an operable decision workflow for risk assessment. Although using a single tree can reduce complexity, it also increases the likelihood of less robust predictions. Part of the strength of XGB (and GBM) models is that they iteratively correct the mistakes of previous trees. A process which is lost if only a single tree is used.

\cite{branstad2023gradient} conducted a study with the objective of 
evaluating the performance of GBM models to predict the presence of aflatoxins in corn at two risk thresholds, that is $20 ppb$ and $5 ppb$. These cut off values were chosen based on the U.S. Food and Drug Administration's (FDA's) action level for corn ($20 ppb$) \citep{FDA}, whereas the lower cut off is based on the European standard of $5 ppb$ \citep{EFSA}. Additionally, the authors performed feature engineering, which is the process of transforming raw data into meaningful and informative features with the intention of enhancing the performance of ML algorithms \citep{zheng2018feature}. The data used was historical climate, soil, and aflatoxin data, collected in several sites in Iowa in the years 2010, 2011, 2012, and 2021. As the data had many missing values, the authors use an imputation method, however they note that data from the months of  January, February, and December had to be excluded from the model as there were too many missing values to accurately impute the data. The authors report that the GBM model performed well, achieving high accuracy rates of 96.8\% for the $20 ppb$ threshold and 90.3\% for the $5 ppb$ threshold. The study highlighted the significant influence of the vegetative index (which is a quantitative measure that uses satellite imagery to assess the amount and health of plant life in a specific area) in August on aflatoxins risk for both thresholds, indicating the critical environmental and ecological impact of drought conditions during this month. Additionally, predictors related to soil properties (such as hydraulic conductivity, pH, and bulk density) were found to potentially affect aflatoxin contamination levels before harvest.

\subsubsection{GB for Spectral Data}

\cite{chavez2022single} conducted a study on aflatoxin and fumonisin contamination in single kernel corn. They argue that bulk sampling of the corn may not produce accurate results, and thus focus solely on single kernels. In their study, they performed measurements to show the skewness of the data, as well as calculating weighted sums of toxin contamination. Additionally, they aim to improve single kernel classification performance through the use of different ML applications. Their methodology was to take corn kernels that are already contaminated and scan them using the NIR technique. The samples were then ground and measured for both toxins using the ELISA method (discussed in  \ref{sec:intro} Section). In their work, they used five different ML models to classify both mycotoxins. They are: GBM; RF; least absolute shrinkage and selection operator (LASSO); elastic-net regularised generalised linear models (GLMNET); and support vector machines (SVM). They additionally applied ML algorithms for classifying each individual mycotoxin. For aflatoxin they used: bagged AdaBoost; linear discriminant analysis (LDA); and penalised logistic regression (PLR). For fumonisin classification, GBM and penalised discriminant analysis (PDA) were used. For aflatoxin, they found that GBM was the best performing model, with an accuracy of 83\%, on both the training and the test data. For fumonisin, the PDA model performed the best with an accuracy of 86\% on the test data. However, the authors note that for future studies, opportunities for better classification exist, including increasing the proportion of samples so the algorithm can learn the characteristics of contaminated corn kernels better.

\subsubsection{Gradient Boosting Summary}
The application of GBM across various datasets, from spatio-temporal to spectral data, demonstrates their versatility and potential in predicting mycotoxin contamination in agricultural products. While GBM models generally exhibit high accuracy, there are criticisms concerning the robustness of these models when applied with limited trees, as in the case of \cite{xie2022fungi}, or when handling datasets with substantial missing values, as noted by \cite{branstad2023gradient}. The high accuracy rates reported should be examined for potential overfitting or lack of generalisation to broader datasets. \cite{chavez2022single}'s approach to single kernel analysis opens avenues for improved precision in toxin detection, but also indicates the need for larger sample sizes to enhance model learning.


\subsection{\textbf{Support Vector Machines}}

\begin{figure}[H]
    \centering
    \includegraphics[width=0.8\textwidth]{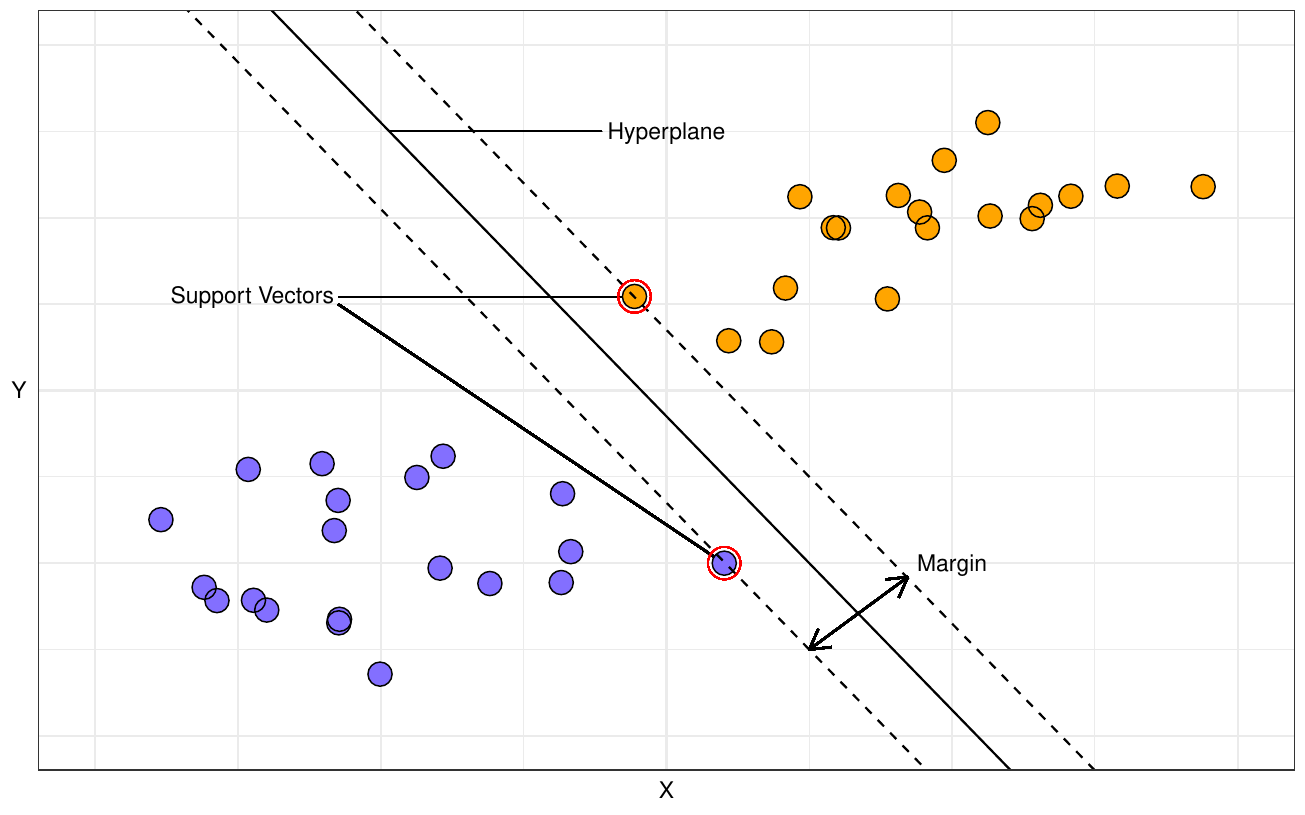}
    \caption{Support Vector Machine Process. The diagram illustrates the SVM's method of finding the optimal hyperplane that maximises the margin between two classes, depicted by the blue and orange points. The support vectors, which are the data points closest to the decision boundary, define the margin.}
    \label{fig:svmProcess}
\end{figure}

Support Vector Machines \citep[SVMs;][]{vapnik1999nature} are a set of supervised learning methods used for classification, regression, and outlier detection. To make predictions, SVMs identify the optimal hyperplane which maximises the margin between the two classes (where the margin is defined as the distance between the nearest data points of each class and the dividing hyperplane). The data points that are closest to the hyperplane and which influence its position and orientation are known as support vectors, as they \textit{support} or define the hyperplane. Figure \ref{fig:svmProcess} illustrates an SVM in action. One of the key advantages of SVMs is their versatility as they can be used on a variety of data types, and are particularly useful for image recognition \citep{burges1998tutorial}. Additionally, they are memory efficient since they only use a subset of training points, called support vectors, in the decision function. However, SVMs require careful tuning of the hyperparameters and an appropriate kernel choice. A kernel is a function used to transform data into a higher dimensional space. By projecting the data into a higher dimension, a kernel makes it possible to find a hyperplane that can effectively separate the classes.
Some of the common kernels include \citep{cristianini2000introduction}:
\begin{enumerate}
 \itemsep0em  
    \item Linear: No nonlinear transformation, suitable for linearly separable data.
    \item Polynomial: Suitable for non-linearly separable data; involves higher degree terms of the features.
    \item Radial Basis Function: Good for non-linear data; uses a Gaussian distribution.
    \item Sigmoid: Similar to the sigmoid function in logistic regression.
\end{enumerate}
Additional hyperparameters include: (i) Gamma: needed for all kernels except linear. It determines the extent of the influence that a single training example has. Low values indicate a wide reach, and high values indicate a close reach. A high gamma value can cause the model to overfit.  (ii) Degree: This is only relevant for polynomial kernel. It defines the degree of the polynomial used in the kernel. A higher degree can model more complex relationships but increases the risk of overfitting. (iii) Coef0: This is a parameter for polynomial and sigmoid kernels that adjusts the independent term in the kernel function. It is often called the kernel bias.

\subsubsection{SVMs for Spectral Data}
In the review of the literature concerning the use of SVMs in mycotoxin detection, it was found that they were overwhelmingly used for image recognition and as such, primarily use spectral data. For example, \cite{almoujahed2022detection} use several ML models (SVM, NN, and LR) for the classification of Fusarium head blight in wheat, using spectral data. The data was collected in the years 2020 to 2021 at a single site in Belgium, with the experiment using eight varieties of wheat.  The found that the SVM model outperformed both the NN and LR method in classifying contaminated wheat in every variety, with a classification accuracy of 96.5\% on the test data (with NN and LR achieving an accuracy of 82.9\% and 82.5\%, respectively).

In a similar study, \cite{kim2023rapid} use three different imaging methods alongside ML classification models to test ground corn samples for the presence of aflatoxin and fumonisin, both as individual contaminants and in combination. Two classification models were used, partial least squares–discriminant analysis (PLS-DA) and SVM, using specific threshold values for each mycotoxin. The naturally contaminated corn samples were obtained  from the Office of Texas State Chemist, which in turn collected the samples from different feed companies located around Texas. They found that the SVM performed better than the PLS-DA with a classification accuracy of 89.1\%, 71.7\%, and 95.7\%, for each imaging technique. The imaging method with the highest accuracy was short-wave infrared (SWIR) method. 

In a study concerning the detection of \textit{Aspergillus parasiticus} in corn kernels using NIR hyperspectral imaging, conducted by \cite{zhao2017early}, the authors use SVMs to compare the performance of multiple different preprocessing and imaging techniques. For their study, corn kernels were harvested from Hefei City, Anhui Province, China in 2015. Each day (for a period 7 days), 36 sterilised corn kernels were inoculated with \textit{Aspergillus parasiticus} and were grouped into four groups depending on the day of inoculation. From this an SVM was used to determine which groups were infected using different preprocessing techniques. Additionally, this study examined the orientation of the kernel in the image to determine if this property had an effect on predictive performance. They found that the best preprocessing method was a combination of the standard normal variate (SNV) and moving average smoothing (MAS) methods, with an accuracy of 91.67\% for detecting contaminated kernels using the validation data. They also found that the performance of the classified models was influenced by orientation; however, the models built using data from a mix of kernels with their germs facing both up and down still achieved an accuracy of 84.38\% on the validation data. 

\subsubsection{Support Vector Machine Summary}
In the reviewed work, SVMs have demonstrated considerable accuracy in mycotoxin detection through spectral data analysis. However, as with other ML methods reviewed, the consistently high classification accuracy reported raises questions about potential overfitting and the representativeness of the datasets used. Moreover, factors such as kernel orientation (which refers to the way in which the kernel function transforms the input data into a higher-dimensional space to find an optimal boundary between classes) significantly influence SVM performance, indicating that model robustness may be context-dependent. The choice of kernel and its parameters, like orientation, scale, and type, are critical in shaping the decision surface and thus the SVM's ability to generalise from training to unseen data


\subsection{Other ML Methods}
In this section we cover the remaining ML methods. These include Decision Trees and Bayesian Networks and have been grouped together as they make up a minority of the reviewed work. As such, they are not separated by the type of data used and all data types are discussed together. 

\subsubsection{\textbf{Decision Trees}}
\label{sec:DTs}

Decision tree (DT) learning is a type of non-parametric supervised learning algorithm used for both classification and regression tasks \citep{quinlan1986induction, breiman1984classification}. A DT is a flowchart-like structure, resembling a tree structure with branches representing decision paths and leaves (or terminal nodes) representing predicted outcomes (see Figure \ref{fig:decisionTreeSingle} in Section \ref{sec:RF}). A DT splits the data into subsets based on the value of input features. Splits are chosen to maximise the separation of the classes, based on measures like Gini impurity or information gain \citep{breiman1984classification}. This process continues recursively until a stopping criterion is met, resulting in a tree where each path represents a decision pathway that leads to a predicted outcome. The advantages of decision trees include their simplicity, interpretability, and ability to handle both numerical and categorical data. However, DTs have a tendency to overfit, especially when a tree is particularly deep \citep{breiman1984classification}. This can be mitigated by pruning the tree or setting a maximum depth of the tree via the use of hyperparameters. As this method is a tree based approach, there is an overlap with RF and GB, in terms of hyperparameters. Some of these include; maximum depth, minimum samples split, and minimum samples leaf (i.e., the minimum number of samples needed to be at a leaf node. Setting this parameter can ensure that each leaf node represents a reasonable number of samples, which can smooth the model, particularly for regression tasks, and prevent overfitting.)

The use of DTs in the field of mycotoxin detection is quite varied. For example, in a study conducted by \cite{kos2016novel}, in which they assess the use of an electronic nose to identify DON contamination of wheat samples, an extension of decision trees called Classification And Regression Trees (CART) \citep{breiman1984classification} was used to classify samples based on four thresholds of DON contamination (1750, 1250, 750, and 500 $\mu g/kg$). For this study, 214 wheat samples were collected from northern Italy during the years 2014–2015 and 2017–2018. For the threshold values of $\ge$ 1250 $\mu g/kg$, the accuracy of sample classification was the highest, ranging between 88\% and 92\%. The lower thresholds of $\le$ 750 $\mu g/kg$ were found to be the least accurate, with an accuracy of $<$ 83\%. The authors proposed that the reduced sensitivity of the instrument at lower DON concentrations might explain this drop in accuracy.

\cite{kos2016novel} examined the classification of DON mycotoxin-contaminated corn and peanuts at regulatory limits using spectral data. The spectral data were analysed using a bootstrap-aggregated (bagged) DT approach, focusing on the protein and carbohydrate absorption bands of the spectrum. The corn samples were obtained by Saatbau Linz (Linz, Austria) and the Cereal Research Centre (Szeged, Hungary). For the peanuts, 92 different, infected samples were purchased from public markets in Tanzania, Mozambique and Burkina Faso. The authors demonstrated that the DT method could classify corn samples at the 1750 and 500 $\mu g/kg$ thresholds for DON with an accuracy of 79\% and 85\%, respectively. Additionally, it was able to classify peanut samples for aflatoxin at 8 $\mu g/kg$ with a 77\% accuracy.

In a study related to identifying and predicting risks related to the presence of fumonisins in breakfast cereal products, \cite{purchase2023association} developed a model specifically designed to predict the risk of fumonisins contamination, with a particular emphasis on a mixture of ingredients. In their research, fifty-eight distinct breakfast products were purchased from local grocery stores in Florence, Italy, during 2019. The selection criteria for purchasing breakfast products included: (i) products with packaging sizes ranging from 200 to 500 grams, including both plastic and non-plastic materials; (ii) items sourced from retail shops; and (iii) products primarily made of wheat, corn, dry fruits, rice, and oats. Principal Component Analysis (PCA) and k-means clustering were employed to explore the connection between cereal ingredients, their composition and packaging, and the concentration of fumonisins. Findings suggested that the fumonisins concentration might be linked to complex, non-linear interactions among various factor variables. To explore this potential and identify the factors most closely linked with high concentrations, DTs were employed. Two Decision Trees (DTs) were developed, with the first indicating a relationship between high concentrations of fumonisins and cereal products rich in corn, particularly when combined with high levels of sodium or rice. The second tree highlighted a link between corn and either high sodium or high-fat concentrations. In both models, the presence of plastic packaging appeared to mitigate the concentration of fumonisins to a certain degree.

\subsubsection{\textbf{Bayesian Network}}

Bayesian networks (BN) are a type of probabilistic graphical model that use Bayesian statistics to represent and infer the conditional dependencies between different variables in a dataset \citep{jensen2007bayesian}. The networks are structured as a directed acyclic graph (DAG), with feature nodes representing variables and edges indicating probabilistic relationships between them. Predictions in BNs are made through a process called probabilistic inference, which involves calculating the likelihood of certain outcomes based on known information and the network's structure. In contrast to linear regression models, BN models excel at analysing variable dependencies, handling non-linear interactions, and incorporating diverse types of data \citep{buritica2015consequence}. The strengths of BN include the handling of uncertainty, the integration of prior knowledge with observed data (thereby enhancing the model's predictive capabilities), and interpretability. However, some disadvantages of using BNs exist. As the number of variables increases, the complexity of the network and the computational resources required for inference can grow exponentially.

In a study aimed for predicting DON contamination in wheat, \cite{liu2018comparison} compare three different modelling approaches. These are: a mixed effect LR method;
a mechanistic model (which simulates the mechanisms of plant and fungus development stages and their interactions) adapted to the current data; and a BN. These were all used to predict DON contamination. The data used was collected in the Netherlands over the years 2001 to 2013. The results of the experiments showed that all three models performed well, with the LR method performing the best, achieving an accuracy of 88\% for detecting DON contamination. However, the authors note that this model is greatly reliant on both the specific location and the available data, and it requires that all input data be present. The mechanistic model achieved an accuracy of 80\%, while the BN achieved an 86\% accuracy. However, the authors note that the BN is easier to implement when the data is incomplete, when compared to the other methods.

\cite{guo2020dynamic} construct transcriptional regulatory networks (TRNs) using a BN algorithm called the \textit{module network} algorithm. TRNs are complex systems in biology that describe the relationships and interactions between various proteins and genes involved in the process of \textit{transcription} \citep{babu2004structure}, where transcription is the process by which the information encoded in a section of DNA is transcribed to produce a complementary RNA strand. The goal of their work was to understand how specific gene groups (modules) in the fungus \textit{Fusarium graminearum} regulate biological processes.
The authors report that their network inference is of high credibility, with 81.8\% of the evaluable modules classified as high or moderate confidence based on their validation against a variety of evidence sources. This suggests a robust alignment of the inferred network with the existing understanding of the biological processes within \textit{Fusarium graminearum}.

\subsubsection{Other ML Methods Summary}

Decision Trees have shown varying degrees of effectiveness in detecting mycotoxins, as evidenced by diverse research outcomes. The use of CART to classify contaminated wheat samples achieved higher accuracy at certain thresholds but showing diminished performance at lower contamination levels. A bagged DT approach showed moderate success, suggesting that while DTs are capable classifiers, their accuracy can vary significantly based on the mycotoxin levels and sample types. The  application of these methods include potential issues with model sensitivity, particularly at lower toxin concentrations, and a reliance on the quality of the data. These factors underscore the need for careful calibration and validation of DTs in diverse settings for reliable mycotoxin detection.

BNs have shown effectiveness in mycotoxin detection, as demonstrated in various studies, but with some limitations. \cite{liu2018comparison} compared BNs with other models for predicting DON contamination in wheat, achieving a respectable 86\% accuracy. However, they highlighted BNs' advantage in handling incomplete data, a significant benefit over other methods like logistic regression. The reviewed applications show BNs' flexibility and efficiency, though their performance can be contingent on data completeness and specific biological contexts, which may limit their broader applicability.


\section{Conclusion}
\label{sec:conclusion}

Our research focuses on highlighting and evaluating different ML models for monitoring and predicting the presence of mycotoxins in common crops. We conducted an extensive literature review of over 45 studies performed within the years 2013 to 2023. The number of publications in each field has grown significantly over the ten years reviewed, however the application of ML in the area of monitoring and predicting mycotoxins is still in its infancy and despite the promise of ML methods in mycotoxin detection, their adoption in industry has been cautious. This is likely due to the high operational costs associated with advanced techniques like hyperspectral imaging, as opposed to the use of ML methods themselves. The prevalence of such data-intensive methods raises questions about the feasibility of widespread implementation, particularly in resource constrained settings.

We found that the most common data type was spectral or image data, and as such the most common ML method used was NNs, as they can be readily applied to image data. RFs were the second most popular ML method, and have gained traction due to their robustness and ease of implementation. Additionally, most of the studies reviewed used classification ML techniques to distinguish contaminated from healthy crops. The high predictive accuracy reported in the reviewed studies suggests that these methods represent a promising approach for mycotoxin detection and enhancing food safety in general. However, a point to note is that the reported high accuracy of the ML model's predictions, often exceeding 90\%, may not fully account for the homogeneity of training and test sets within individual laboratories. This homogeneity can result in overfitting, where models appear highly accurate in a controlled setting but may not perform as well under the variable conditions of real-world applications.

A critical observation from this review is that the lack of detailed hyperparameter descriptions further complicates the landscape, as these parameters are crucial for the replication and validation of ML models. Without clear reporting on hyperparameter tuning, the ability to reproduce results and validate findings becomes challenging, hindering the progression towards robust and reliable ML applications in food safety. The majority of the reviewed studies do not provide open access to code and many have limited access to data, further impeding the reproducibility of the described methods.  As the field matures, there is a need for standardisation in reporting practices and for developing models that can reliably perform across diverse laboratory conditions and datasets.

Although this work focused on the application of the most popular ML methods, numerous other ML and statistical techniques have been applied to mycotoxin detection data. For example, in a study by \cite{de2019rapid}, classification models such as Partial-Least Squares-Discriminant Analysis (PLS-DA) and Principal Component-Linear Discriminant Analysis (PC-LDA) were employed to distinguish between wheat samples with high and low contamination. Additionally, statistical techniques like PCA are often used as a dimension reduction method. \cite{shen2019line, jha2021rapid}, and \cite{milicevic2019impact} all use PCA when dealing with high-dimensional image data. 

As the field is growing, there are numerous avenues for future work. More extensive research could be conducted that directly compares different ML models under a standardised set of hyperparameters. This will provide clearer insights into the most effective techniques in specific contexts related to mycotoxin detection. Another avenue in 
model interpretability. Given the critical nature of food safety, future research could also focus on improving the interpretability of ML models. Techniques like SHAP (SHapley Additive exPlanations) \citep{shapley1953value} and LIME (Local Interpretable Model-agnostic Explanations) \citep{ribeiro2016should} can be used to make the models' decisions more transparent and trustworthy.

\vfill
\noindent
\footnotesize
\textbf{Acknowledgements:} 
This work was done as part of the Mycotox-I project that is kindly supported by the Department of Agriculture, Food and the Marine (DAFM) and The Department of Agriculture, Environment and Rural Affairs (DAERA), grant number 2021R460. Andrew Parnell’s work was supported by the SFI Centre for Research Training in Foundations of Data Science 18/CRT/6049, and SFI Research Centre award 12/RC/2289\_P2. For the purpose of Open Access, the author has applied a CC BY public copyright licence to any Author Accepted Manuscript version arising from this submission

\newpage
\section{Appendix}
\label{sec:appendix}
In Table \ref{tab:abbrev} we provide a list of the used abbreviations and nomenclature used in this work.

\begin{table}[H]
    \centering
    \renewcommand{\arraystretch}{0.8} 
    \begin{tabular}{c|c}
    \hline
        \textbf{Abbreviation} & \textbf{Meaning} \\
        \hline
         NN & Neural Network \\
         CNN & Convoluted Neural Network \\
         RF &  Random Forest \\
         GBM & Gradient Boosted Machine \\
         XGB & eXtreme Gradient Boosted Machine \\
         DT & Decision Trees \\
         CART & Classification And Regression Trees \\
         SVM & Support Vector Machine \\
         BM & Bayesian Models \\
         BN & Bayesian Network \\
         LDA & Linear Discriminant Analysis \\
         PDA & penalised discriminant analysis \\
         LReg & Linear Regression \\
         LR & Logistic Regression \\
         MLR & Multiple Linear Regression \\
         LASSO & Least Absolute Shrinkage and Selection Operator \\
         GLMNET & elastic-net regularised generalised linear models \\
         PLS-DA & Partial Least Squares–Discriminant Analysis \\
         sPLS-DA & sparse Partial Least Square Determination Analysis \\
         PCA & Principal Components Analysis \\
         MLP & Multi-Layer Perceptron \\
         BLUP & Best Linear Unbiased Prediction \\
         PCC & Pearson Correlation Coefficient \\
         RMSE & Root Mean Square Error \\
         $R^2$ & Coefficient of Determination \\
         AUC & Area Under the Curve \\
         NIR & Near-Infrared Spectroscopy \\
         DON & Deoxynivalenol \\
    \end{tabular}
    \caption{Nomenclature used in our paper.}
    \label{tab:abbrev}
\end{table}

\newpage
\bibliographystyle{asa}

\bibliography{myc.bib}

\end{document}